pdfoutput=1                                                            
\documentclass{aa}  
\usepackage{graphicx}
\usepackage{txfonts}
\usepackage[colorlinks,citecolor=blue]{hyperref}
\begin{document}

   \title{The magnetic drivers of campfires seen by the Polarimetric and Helioseismic Imager (PHI) on Solar Orbiter}

  \author{F. Kahil\inst{1}\thanks{\hbox{Corresponding author: F. K.} \hbox{\email{kahil@mps.mpg.de}}}
     \and
   J.~Hirzberger\inst{1} \and
   S.K.~Solanki\inst{1,10} \and
   L.~P.~Chitta\inst{1} \and
   H.~Peter\inst{1} \and
   F.~Auch\`ere\inst{3} \and
   J.~Sinjan \inst{1} \and
   D.~Orozco Su\' arez\inst{2} \and
   K.~Albert\inst{1} \and
   N. Albelo Jorge\inst{1} \and
   T.~Appourchaux\inst{3} \and 
   A.~Alvarez-Herrero\inst{4} \and
   J.~Blanco Rodr\'iguez\inst{5} \and
   A.~Gandorfer\inst{1} \and
   D.~Germerott\inst{1} \and
   L.~Guerrero\inst{1} \and
   P.~Guti\'errez M\'arquez\inst{1} \and
   M.~Kolleck\inst{1} \and
   J.C.~del Toro Iniesta\inst{2} \and
   R.~Volkmer\inst{6} \and
   J.~Woch\inst{1} \and 
   B.~Fiethe\inst{7} \and
   J.M.~Gómez Cama\inst{8} \and
   I.~P\' erez-Grande\inst{9} \and 
   E.~Sanchis Kilders\inst{5} \and
   M.~Balaguer Jiménez\inst{2}~\and
   L.R.~Bellot Rubio\inst{2} \and
   D.~Calchetti\inst{1} \and
   M.~Carmona\inst{8} \and
   W.~Deutsch\inst{1} \and 
   G.~Fern\'andez-Rico\inst{1,9} \and
   A.~Fern\' andez-Medina\inst{4} \and
   P.~Garc\'\i a Parejo\inst{4} \and 
   J.L.~Gasent-Blesa\inst{5} \and 
   L.~Gizon\inst{1,11} \and
   B.~Grauf\inst{1} \and 
   K.~Heerlein\inst{1} \and
   A.~Lagg\inst{1} \and
   T.~Lange\inst{7} \and 
   A.~L\' opez Jim\' enez\inst{2} \and 
   T.~Maue\inst{6} \and 
   R.~Meller\inst{1} \and
   H.~Michalik\inst{7} \and
   A.~Moreno Vacas\inst{2} \and
   R.~M\" uller\inst{1} \and
   E.~Nakai\inst{6} \and 
   W.~Schmidt\inst{6} \and
   J.~Schou\inst{1} \and
   U.~Sch\"uhle\inst{1}\and
   J.~Staub\inst{1} \and
   H.~Strecker \inst{2} \and 
   I.~Torralbo\inst{9} \and 
   G.~Valori\inst{1} \and
   R.~Aznar Cuadrado\inst{1} \and
   L.~Teriaca\inst{1} \and
   D.~Berghmans\inst{12} \and
   C.~Verbeeck \inst{12} \and
   E.~Kraaikamp \inst{12}\and
   S.~Gissot \inst{12}
   }

   \institute{
         Max-Planck-Institut f\"ur Sonnensystemforschung, Justus-von-Liebig-Weg 3,
         37077 G\"ottingen, Germany \\ \email{kahil@mps.mpg.de}
         \and
         Instituto de Astrofísica de Andalucía (IAA-CSIC), Apartado de Correos 3004,
         E-18080 Granada, Spain 
         \and
         Univ. Paris-Saclay, Institut d’Astrophysique Spatiale, UMR 8617,
         CNRS, B\^ atiment 121, 91405 Orsay Cedex, France
         \and
         Instituto Nacional de T\' ecnica Aeroespacial, Carretera de
         Ajalvir, km 4, E-28850 Torrej\' on de Ardoz, Spain
         \and
         Universitat de Val\`encia, Catedr\'atico Jos\'e Beltr\'an 2, E-46980 Paterna-Valencia, Spain
         \and
         Leibniz-Institut für Sonnenphysik, Sch\" oneckstr. 6, D-79104 Freiburg, Germany
         \and
         Institut f\"ur Datentechnik und Kommunikationsnetze der TU
         Braunschweig, Hans-Sommer-Str. 66, 38106 Braunschweig,
         Germany
         \and
         University of Barcelona, Department of Electronics, Carrer de Mart\'\i\ i Franqu\`es, 1 - 11, 08028 Barcelona, Spain
         \and
         Instituto Universitario "Ignacio da Riva", Universidad Polit\'ecnica de Madrid, IDR/UPM, Plaza Cardenal Cisneros 3, E-28040 Madrid, Spain
         \and
         School of Space Research, Kyung Hee University, Yongin, Gyeonggi-Do, 446-701, Korea
         \and
         Institut f\"ur Astrophysik, Georg-August-Universit\"at G\"ottingen, Friedrich-Hund-Platz 1, 37077 G\"ottingen, Germany
        \and
         Solar-Terrestrial Centre of Excellence-SIDC, Royal Observatory of Belgium, Ringlaan-3-Av. Circulaire, 1180 Brussels, Belgium}

   \date{Received ; accepted }

 
  \abstract
   {The Extreme Ultraviolet Imager (EUI) on board the Solar Orbiter (SO) spacecraft observed small extreme ultraviolet (EUV) bursts, termed campfires, that have been proposed to be brightenings near the apexes of low-lying loops in the quiet-Sun atmosphere. The underlying magnetic processes driving these campfires are not understood. }
   {During the cruise phase of SO and at a distance of 0.523\,AU from the Sun, the Polarimetric and Helioseismic Imager on Solar Orbiter (SO/PHI) observed a quiet-Sun region jointly with SO/EUI, offering the possibility to investigate the surface magnetic field dynamics underlying campfires at a spatial resolution of about 380~km.}
   {We used co-spatial and co-temporal data of the quiet-Sun network at disc centre acquired with the High Resolution Imager of SO/EUI at 17.4 nm (HRI$_{\rm EUV}$, cadence 2\,s) and the High Resolution Telescope of SO/PHI at 617.3 nm (HRT, cadence 2.5\,min). Campfires that are within the SO/PHI$-$SO/EUI common field of view were isolated and categorised  according to the underlying magnetic activity.}
   {In 71\% of the 38 isolated events, campfires are confined between bipolar magnetic features, which seem to exhibit signatures of magnetic flux cancellation. The flux cancellation occurs either between the two main footpoints, or between one of the footpoints of the loop housing the campfire and a nearby opposite polarity patch. In one particularly clear-cut case, we detected the emergence of a small-scale magnetic loop in the internetwork followed soon afterwards by a campfire brightening adjacent to the location of the linear polarisation signal in the photosphere, that is to say near where the apex of the emerging loop lays. The rest of the events were observed over small scattered magnetic features, which could not be identified as magnetic footpoints of the campfire hosting loops.}
   {The majority of campfires could be driven by magnetic reconnection triggered at the footpoints, similar to the physical processes occurring in the burst-like EUV events discussed in the literature. About a quarter of all analysed campfires, however, are not associated to such magnetic activity in the photosphere, which implies that other heating mechanisms are energising these small-scale EUV brightenings.  }


   \maketitle
%
\section{Introduction}

During the commissioning phase of Solar Orbiter \citep[SO,][]{muller_solar_2020} on May 30, 2020, small-scale brightenings were seen in the extreme ultraviolet (EUV) images (at 17.4 nm) obtained with the High Resolution Imager (HRI$_{\rm EUV}$) of the Extreme Ultraviolet Imager \citep[EUI;][]{rochus_solar_2020}. These small transient brightenings were dubbed campfires (for their flame-like appearance) and their properties have been recently reported in \cite{berghmans_extreme-uv_2021}. 

Generally, transient events observed in the solar atmosphere are categorised based on their formation heights and magnetic environment, and they are studied for their role in atmospheric heating. 
Ultraviolet bursts are a common small-scale phenomenon. They are observed in regions of emerging magnetic flux, in active regions and in the quiet Sun. They are associated with an enhancement of UV and EUV emission and peculiarities in their spectral profiles \citep{young_solar_2018}. These include Ellerman bombs in the photosphere \citep{1917ApJ....46..298E}, their counterparts in the chromosphere \citep{2014Sci...346C.315P}, and explosive events in the transition region \citep{dere_explosive_1989}. 
In the lower corona, coronal bright points (CBPs) are bright in EUV and X-rays and correspond to small-scale loops. They have lifetimes of several minutes to hours and have sizes larger than 10 Mm \citep{madjarska_coronal_2019}. They are generally larger than the aforementioned transients.
Extreme UV brightenings have been observed over a wide range of energies and spatial scales, generally exhibiting power-law-like distributions of their energy content \cite[e.g.][]{aschwanden_nanoflare_2002}, and it is still not clear whether they might contribute a sufficient amount of energy input to heat the upper solar atmosphere \citep{chitta_extreme-ultraviolet_2021}.

SO/EUI now pushes the boundaries to smaller scales when recording events in quiet-Sun EUV emission. 
The smallest of the campfires detected with the EUI instrument are the smallest ever detected transient events seen in EUV radiation \cite[][]{berghmans_extreme-uv_2021}. Using triangulation combining EUI and the Atmospheric Imaging Assembly \cite[AIA;][]{lemen_atmospheric_2012} on the Solar Dynamics Observatory \cite[SDO;][]{pesnell_solar_2012}, it became apparent that campfires are formed in a height range of only $1000$ to $5000$~km above the photosphere \cite[][]{zhukov_stereoscopy_2021}. Thus they barely stick out of the chromosphere.
\cite{berghmans_extreme-uv_2021} found the campfires are mostly associated with the chromospheric network, as seen in the co-spatial and co-temporal images in the Lyman-Alpha channel (HRI$_{\rm Lya}$) of HRI at 121.6 nm.
The chromospheric network is associated with strong magnetic concentrations \citep[e.g.][]{kahil_brightness_2017, barczynski_emission_2018}, which suggests that the campfires are magnetic in origin. 

 A direct relation of the campfires to the photospheric magnetic field has not been completely established. \cite{zhukov_stereoscopy_2021} looked into the underlying magnetic field in the Helioseismic Magnetic Imager \cite[HMI;][]{schou_design_2012} full-disc magnetograms on board SDO and found that some of them belong to magnetic bipolar structures and others are projected onto areas of a weak magnetic field. They suspect that the lower resolution of HMI (1{\arcsec}) could be the reason why the footpoints of some of the campfires could not be resolved \citep[see also the discussion related to magnetic fields underlying EUV bursts observed with AIA in][]{chitta_extreme-ultraviolet_2021}.
 
 \cite{panesar_magnetic_2021} extended the study of \cite{zhukov_stereoscopy_2021} of the magnetic components of campfires using the HMI full disc magnetograms. They found that the majority of campfires are accompanied by cool plasma structures with magnetic flux cancellation events (with a rate of $10^{18}$ Mx/hour). They propose that flux cancellation triggers the cool-plasma eruption (as seen in the AIA images) and that this gives rise to campfires.
 
In an attempt to investigate how campfires relate to the magnetic field in the lower photosphere, \citet{chen_transient_2021} used MURaM \citep{vogler_simulations_2005} simulations to model the photosphere, chromosphere, and lower transition region of a quiet-Sun network region. They then synthesized the emission in the working wavelengths of SO/EUI  and degraded their simulations to match the spatial resolution of the SO/EUI  observations. They found that the modelled transient brightenings do not show significant magnetic cancellation or emergence in the corresponding line-of-sight magnetic field maps. Instead, the modelled campfires are associated with component reconnection at coronal heights with varying field-line geometries, consistent with the triangulated height of the campfires. Also, the model matches the observations in a number of other properties, such as the lifetime, size, and aspect ratio.


\begin{figure*}
\centering
     \includegraphics[width=\textwidth]{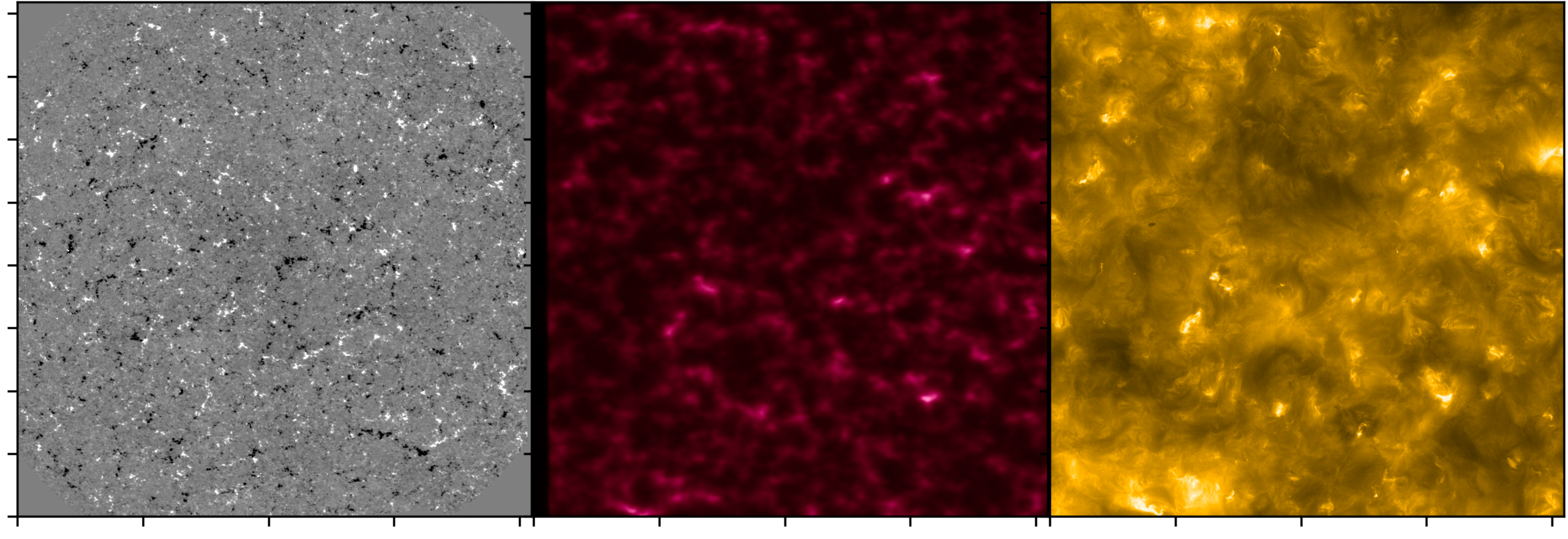}
      \caption{Photospheric magnetic field and images from the chromosphere and corona. From left to right: Inverted $\rm B_{\rm LOS}$ map of SO/PHI in 6173\, {\AA} at 16:59:08 UT saturated at $\pm 40$~G, the nearly co-temporal HRI$_{\rm Lya}$ image in 1216\,{\AA} recorded at 16:58:55 UT, and the HRI$_{\rm EUV}$ image in 174\,{\AA} recorded at 17:13:25 UT. The data were taken during the cruise phase of SO on February 23, 2021. The maps shown here were calibrated, but the SO/PHI magnetogram is not aligned to SO/EUI filtergrams. The three images have a FOV of $1024{\arcsec}\times1024${\arcsec}, which corresponds to about $398\times398$ Mm$^2$ on the Sun at a SO$-$Sun distance of 0.523 AU.}

      \label{data_rscw2}

\end{figure*}

\begin{figure*}
\centering
\includegraphics[width=\textwidth]{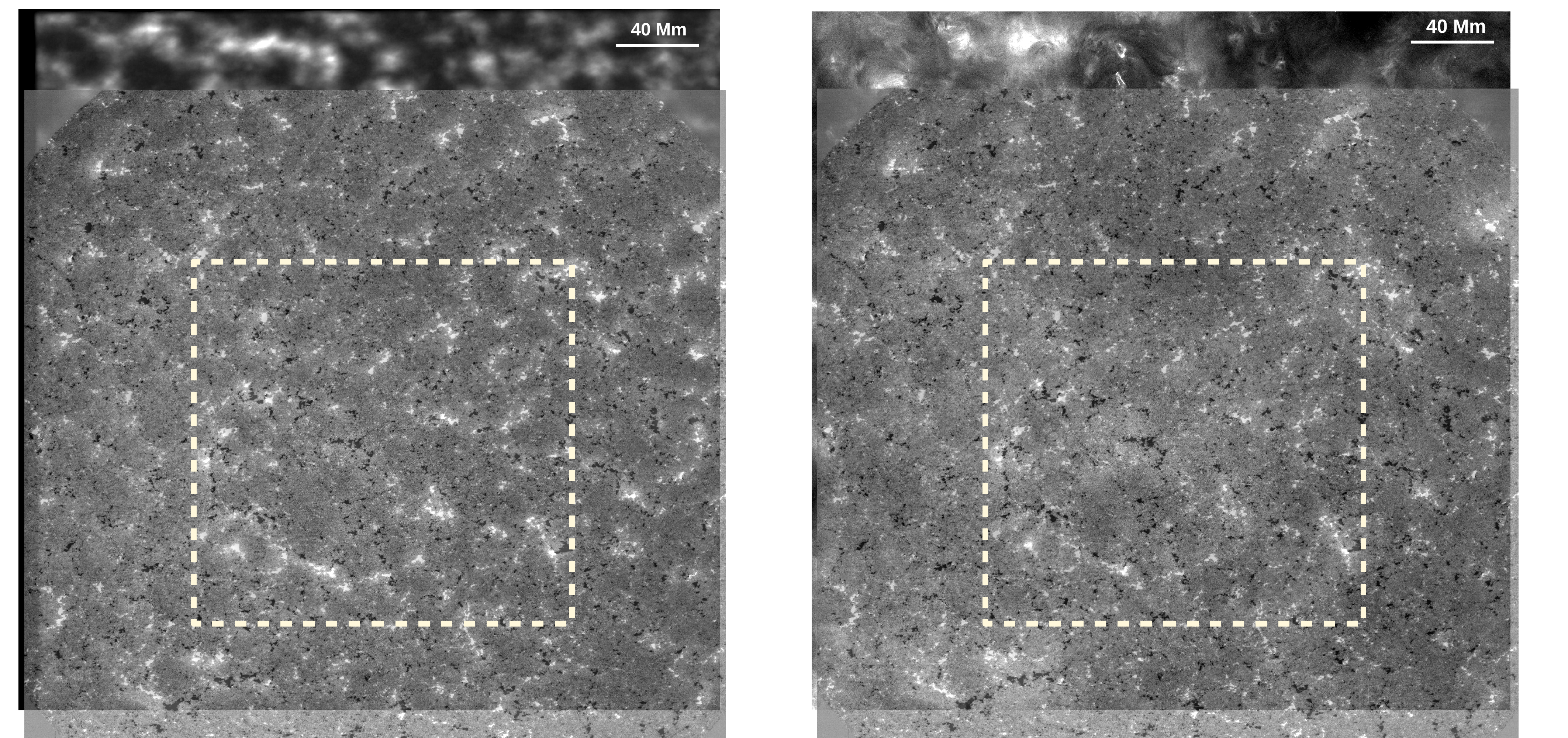}
  \caption{ SO/PHI$-$SO/EUI aligned FOVs. The SO/PHI LOS magnetogram of the first full dataset of $1024{\arcsec}\times1024${\arcsec} is overlaid on the co-aligned HRI$_{\rm Lya}$ (left) and HRI$_{\rm EUV}$ (right). The SO/PHI magnetogram was made partially transparent to make the correlated features between the two instruments more visible. The dashed box in the middle of the SO/PHI FOV is the observed region in the other seven SO/PHI datasets to which the aligned HRI$_{\rm EUV}$ images were cropped. Solar north is up and solar west is to the right.}
  
     \label{hrt_eui_shift}
\end{figure*}

\begin{figure*}
\centering
     \includegraphics[width=\textwidth]{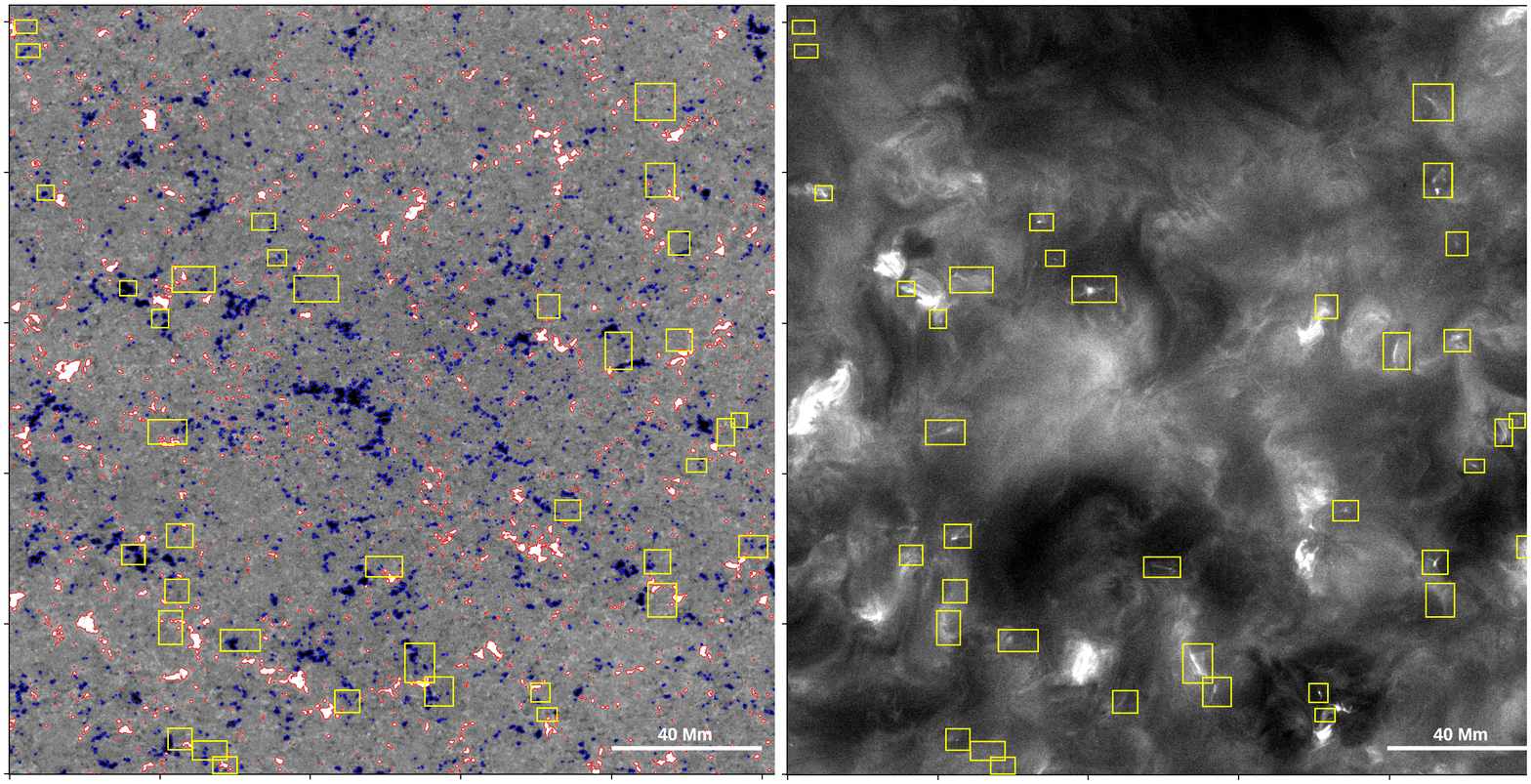}
      \caption{Distribution of campfire events over the SO/PHI LOS magnetogram (left) and co-aligned HRI$_{\rm EUV}$ image (right). The contours on the magnetic field map enclose magnetic features with $\rm B_{\rm LOS}$ above $3\sigma_B$ (21~G). Blue and red contours correspond to negative and positive flux, respectively. The yellow boxes in both figures enclose the studied campfire events which are located approximately in the centre of each box. The LOS magnetogram is saturated at $\pm 40$~G, and the size of the FOV is $512{\arcsec}\times512${\arcsec}.}
      \label{contours_hrt}

\end{figure*}

One of the problems of understanding the physics of small-scale transients in the solar atmosphere is the disparity between the spatial resolution of the EUV observations and of the photospheric magnetic field data. In some cases, high-resolution polarimetric observations of magnetic flux cancellation have to be paired with much coarser coronal observations, for example prohibiting a clear link of flux cancellation to the formation of coronal loops \cite[][]{chitta_solar_2017}. On the other hand, in several cases, the few pieces of data acquired in the EUV at a resolution below 0.5{\arcsec} needed to be paired with comparably low resolution magnetic field data, for example preventing clear magnetic signatures related to miniature coronal loops from being identified \cite[][]{barczynski_miniature_2017}.
Here SO can provide a major step forward by combining the SO/EUI imaging of the upper atmosphere with observations of the photospheric magnetic field recorded by the Polarimetric and Helioseismic Imager \cite[SO/PHI,][]{solanki_polarimetric_2020}. The recorded data of both instruments have nearly the same spatial resolution of 1{\arcsec}. At the perihelion of SO during the nominal mission phase, the spatial resolution of both instruments can reach about 200~km on the Sun corresponding to about 0.3{\arcsec} as seen from the Earth \citep{muller_solar_2020}. 

During the second remote sensing checkout window \citep{zouganelis_solar_2020}, the High Resolution Telescope (HRT) of SO/PHI acquired high resolution time series of the quiet-Sun network jointly with the HRI$_{\rm EUV}$ and HRI$_{\rm Lya}$, offering the first possibility to look into the underlying magnetic field with a spatial resolution of 380~km at 0.523 AU. Despite the short time during which the SO/PHI-HRT and HRI$_{\rm EUV}$ datasets overlapped, we were able to study and follow the dynamics of the magnetic features below the detected campfires. In Section~\ref{observations} we describe the data we employ in this study as well as the alignment procedure between the two remote sensing (RS) instruments, SO/PHI and SO/EUI. In Section~\ref{data-analysis} we describe our approach to study the magnetic counterparts of the campfires and present our findings. In Section~\ref{discussion} we discuss our results and compare them to observations of the photospheric magnetic field below other transient events.

\begin{figure*}[ht!]
\centering
\includegraphics[width=\textwidth]{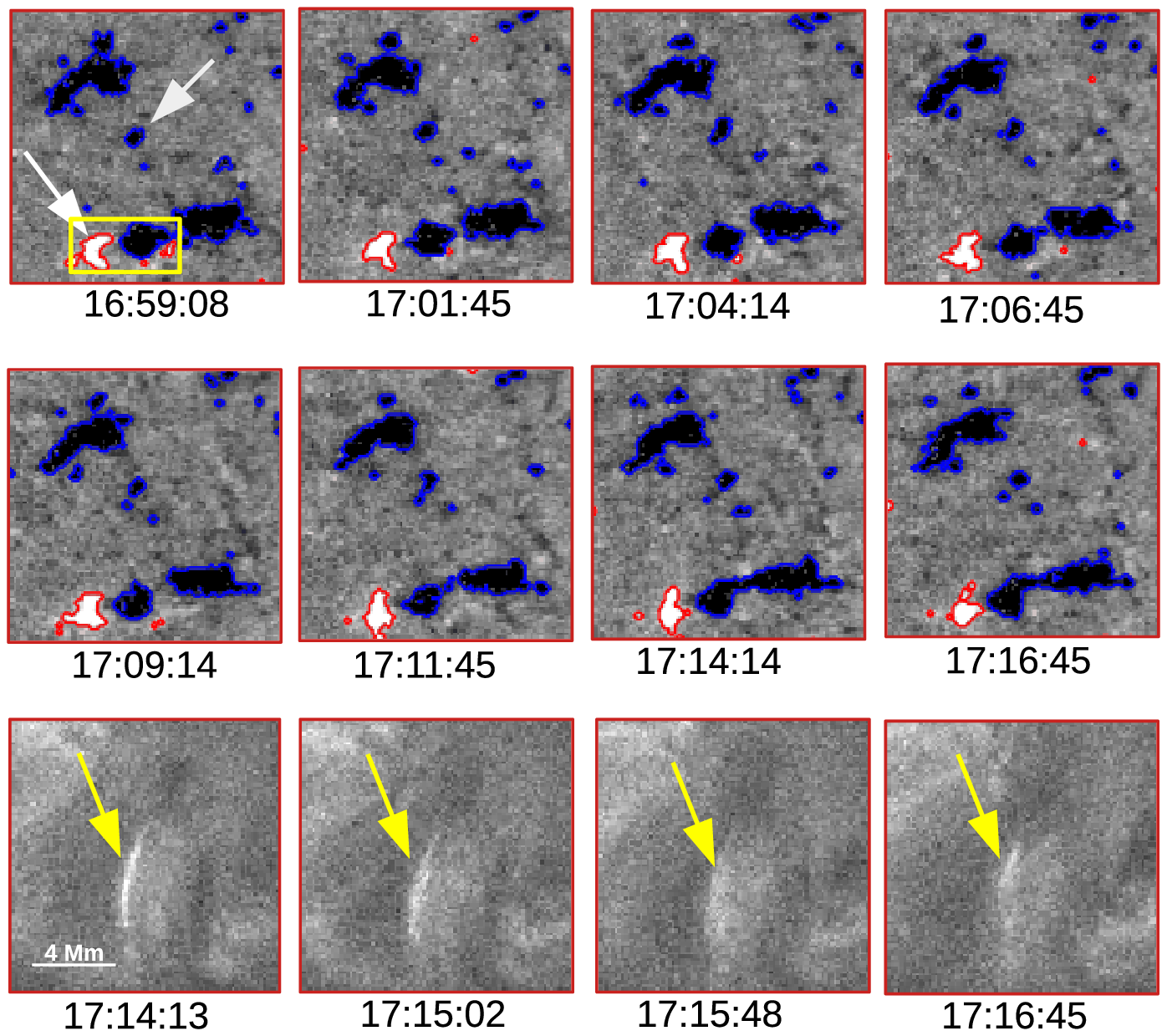}
  \caption{ SO/PHI magnetic field distribution below a campfire event of Category~Ia. The top two rows are the eight subregions ($13\times15$ Mm$^2$) of the SO/PHI magnetograms below the campfire event. The white arrows in the first SO/PHI magnetogram at 16:59:08 UT point to the location of the two main footpoints of the campfire. The yellow box contains the positive main polarity and the opposite polarity feature interacting with it. The last row displays the co-aligned area in HRI$_{\rm EUV}$ for four images acquired within the last two SO/PHI magnetograms (i.e. between 17:14:13 UT and 17:16:45 UT). The yellow arrow in the last row indicates the location of the campfire event which has a loop-like structure. The maximum brightness of this event according to the detection algorithm of \cite{berghmans_extreme-uv_2021} is achieved around 17:15:23 UT. 
          }
     \label{event5a}
\end{figure*}

\begin{figure*}
\centering
\includegraphics[width=\textwidth]{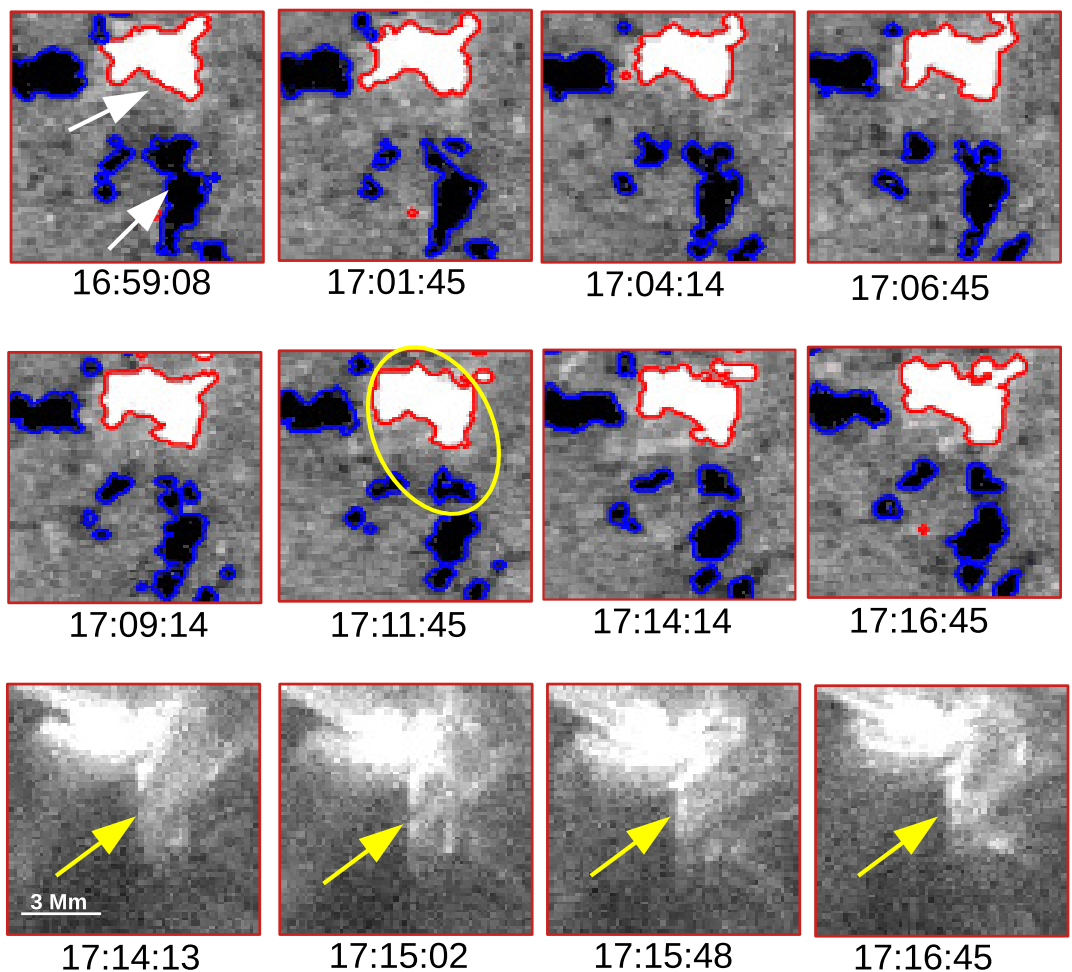}
  \caption{SO/PHI magnetic field distribution below a campfire event of Category~Ib. The main footpoints of the event are indicated by white arrows in the first image in the first row. The yellow ellipse at 17:11:45 UT encloses the smaller feature splitting from the main negative polarity footpoint and approaching the large positive polarity footpoint. In the last row we show the campfire event in HRI$_{\rm EUV}$. The maximum brightness of this event is reached around 17:16 UT. The area just above the campfire event in the HRI$_{\rm EUV}$ images is overexposed. The plotted region has dimensions of $11\times11$ Mm$^2$. 
          }
     \label{event27b}
\end{figure*}

\begin{figure}
\centering
\includegraphics[width=\hsize]{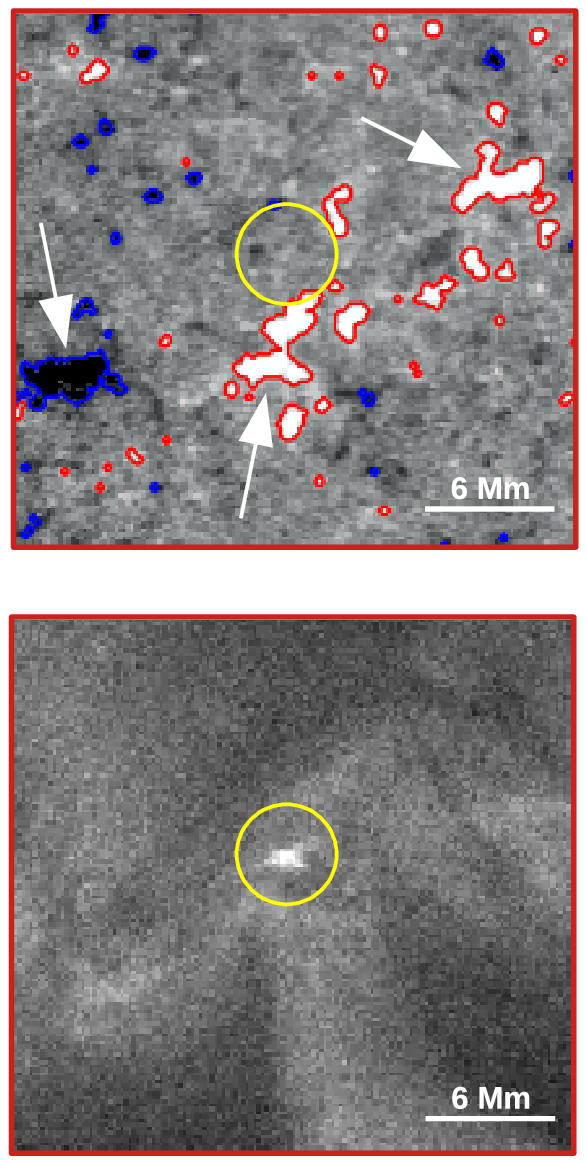}
  \caption{ SO/PHI magnetic field distribution below and around a campfire event of Category II. The yellow circle encloses the detected campfire event in the SO/PHI $\rm B_{\rm LOS}$ (upper panel) and HRI$_{\rm EUV}$ (lower panel) images. The white arrows point to the possible magnetic footpoints of interacting bundles of magnetic field lines at coronal heights, similar to the footpoints' arrangement found by \cite{chen_transient_2021}. The plotted region has dimensions of $25\times25$ Mm$^2$.
          }
     \label{categ2}
\end{figure}

\section{Observations}
\label{observations}

In its cruise phase, SO offered an opportunity for the RS instruments on board to operate simultaneously during the so-called remote sensing checkout windows. During these checkout periods, instruments were tested by acquiring calibration data to ensure they perform well during the nominal mission phase. There was also a possibility to acquire observations of the same solar region to test the accuracy of the pointing and to check for the relative orientation between the different fields of view (FOVs) of the RS instruments. This provided some limited opportunities to acquire first science-grade co-observations of different instruments. The overlap between the FOVs is expected to vary with the different thermal environment and therefore with the distance to the Sun. Techniques to intercalibrate the FOVs between the various instruments are described in \cite{auchere_coordination_2020}. Cross-calibration between images from different instruments with similar formation heights is one of their suggested techniques. In this study, we follow this approach to align the SO/PHI and SO/EUI data and revisit the relative orientation between their FOVs to update it with respect to the co-alignment achieved on the ground. In this work, we are only using the HRT data and so SO/PHI always refers to SO/PHI$-$HRT.
 
On February 23, 2021, the EUI/HRI acquired image sequences between 16:55:15 UT and 16:58:55 UT in Lya (121.6 nm, cadence of 5s) and between 17:13:25 UT and 17:20:59 UT in EUV (17.4 nm, cadence 2s) \citep[see also][for a discussion of small EUV brightenings observed using these 2~s cadence EUV data]{chitta_capturing_2021}.
SO/PHI acquired eight datasets between 16:59:08 UT and 17:16:45 UT, each consisting of 24 images recorded in the Fe\,{\sc i} line at 6173\, \AA{} (four polarisation states $\times$ six wavelengths). A full SO/PHI dataset needed 2.5~minutes to be fully recorded. Only the first dataset has a full FOV of $2048\times2048$ pixels. The remaining seven datasets were cropped to $1024\times1024$ pixels around the central region to reduce telemetry needs. The raw SO/PHI data were dark subtracted, flat-fielded, and demodulated. The Stokes vector was then inverted with the CMILOS code \citep{suarez_usefulness_2007} to get the longitudinal magnetic field ($\rm B_{\rm LOS}$) maps\footnote{The low signal level in the observed quiet-Sun region made the retrieved transverse field very noisy almost everywhere in the FOV.}. 
The angular pixel size achieved by the three high resolution telescopes considered here is about 0.5{\arcsec}, which corresponds to roughly 190~km on the Sun at a distance of 0.523 AU. 

We show in Figure~\ref{data_rscw2} a calibrated SO/PHI magnetogram, as well as level-2 HRI$_{\rm EUV}$ and HRI$_{\rm Lya}$ images \footnote{\url{https://doi.org/10.24414/k1xz-ae04}}. The relative orientation of the cameras between the different telescopes was deduced from the joint analysis of SO and SDO data during the inter-calibration campaign during instrument commissioning on May 30, 2020 at 17:42 UT. At that time, the SO$-$Sun$-$Earth angle was about 30 degrees in solar longitude. We used the SDO/AIA full-disc images in the 30.4 nm filter to compare them with the FOV of the Full-Sun Imager channel of SO/EUI  at 30.4 nm (FSI 30.4). The corresponding full-disc magnetograms of SDO/HMI were used to determine the orientation of the SO/PHI FOV with respect to SDO/HMI, and then via SDO/AIA with respect to EUI/HRI. 

Before aligning both datasets, the SO/PHI data were corrected for geometrical distortion across the camera. Next, the last HRI$_{\rm Lya}$ image (at 16:58:55 UT) was cross-correlated with the first SO/PHI $\rm B_{\rm LOS}$ map (at 16:59:08 UT). The alignment was performed between these two channels because the magnetic network is well visible in both (see Figure~\ref{data_rscw2}). Consequently, the FOV of SO/PHI is found to be shifted southward by about 236 pixels and westward by 10 pixels with respect to the SO/EUI FOV, as shown in Figure~\ref{hrt_eui_shift}. 

Since the rest of the seven SO/PHI datasets are cropped to the central $1024\times1024$ pixels area, we used only that common region in the EUV data to track the magnetic features below campfires. The common HRI$-$HRT registered area is outlined by a box in Figure~\ref{hrt_eui_shift}.

The noise level in the SO/PHI data was estimated by examining the signals in the continuum Stokes~$V$ filtergram and it is equal to $\sigma = 10^{-3}I_c$, with $I_c$ being the mean quiet-Sun intensity. This corresponds to $\sigma_{B}=7$ G in the SO/PHI LOS magnetograms. In addition to the SO/PHI $\rm B_{\rm LOS}$ maps, we used maps of linear polarisation (LP) signals averaged over the SO/PHI line. The total LP (TLP) was computed following the method described by \cite{jafarzadeh_structure_2013}, who applied it to data obtained by the IMaX instrument \citep{martinezpillet_imaging_2011} on the Sunrise balloon-borne observatory \citep{solanki_sunrise:_2010,   barthol_sunrise_2011, gandorfer_filter_2011, berkefeld_wave-front_2011}. The noise in the line-averaged LP was estimated by measuring the TLP in the continuum wavelength and it is equal to $\sigma_{LP} = 7\times 10^{-4} I_c$.

\section{Data analysis and results}
\label{data-analysis}
We used the output of the automated detection algorithm by \cite{berghmans_extreme-uv_2021} to locate the campfires in the HRI$_{\rm EUV}$ data. Their detection was performed on the Carrington projected EUV images, but we used the output coordinates transformed to the image plane of the detector. This was done to avoid any interpolation of the SO/PHI data, which is necessary for this kind of geometrical projection, and to preserve the image quality. We then looked for the events that were within the co-registered region with SO/PHI (dashed box in Figure~\ref{hrt_eui_shift}). 

The HRI$_{\rm EUV}$ timeseries starts 14 minutes after the first SO/PHI dataset, and the SO/PHI time series ends 2.5 minutes after the beginning of the HRI$_{\rm EUV}$ observations. As a result, this study is restricted to the detected events in the co-registered HRI$_{\rm EUV}$ data, which are simultaneous with the last two SO/PHI datasets. This corresponds to about 75 examined HRI$_{\rm EUV}$ images starting at 17:14:13 UT and ending at 17:16:45 UT. For each detected event, we studied the magnetic surroundings from the beginning of the SO/PHI time series to its end. 

The automated campfire detection algorithm of \cite{berghmans_extreme-uv_2021} returned the (x,y) pixel position of the maximum brightness in each time frame. We assigned pixels located within less than 10~Mm from each other and confined within the same magnetic features (as seen in the SO/PHI $\rm B_{\rm LOS}$ maps) to a single event. Accordingly, we end up with 38 well-isolated events and delimit them in yellow in Figure~\ref{contours_hrt}. Some of the events are not visible in the HRI$_{\rm EUV}$ image of Figure~\ref{contours_hrt} because those campfires achieved their peak brightness at a different time than the displayed snapshot.

The distribution of campfires in the SO/PHI LOS magnetogram of Figure~\ref{contours_hrt} indicates that they are associated with the magnetic network outlined by the large magnetic structures. The magnetic network patches in this FOV have a dominant negative polarity.

Magnetic field extrapolations are required to reliably link campfires to the underlying magnetic field patches in order to understand the role of specific magnetic processes (e.g. emergence or cancellation of magnetic flux) in triggering these events. In this first study, however, we relied on a visual inspection to study the magnetic field associated with the selected campfires. We divided the events into two main categories based on the magnetic configuration below them. We describe these two categories in the following subsections. In both of these categories, the campfires are either dot-like events, loop-like events, or events with complex morphologies, which is consistent with the morphology of campfires presented in \cite{berghmans_extreme-uv_2021} (see discussion in Section~\ref{morphology}).

\subsection{Category I: Events projected on magnetic bipolar features }
The majority of the campfires (27 out of 38 events) are associated with magnetic bipolar features, which are the main footpoints of their respective loops. Events belonging to this category should satisfy two criteria. First, the footpoints should be located within a 10 Mm radius centred on the campfire. Second, the footpoints (and magnetic features interacting with the footpoints) should persist throughout the SO/PHI time series and maintain $\rm B_{\rm LOS}$ signals above $3\sigma_B=21$~G. In all of these events, we see signs of magnetic flux cancellation. We further divide the events belonging to this category into two sub-categories.\\

    {\it Category Ia.} In 18 cases, we found an opposite polarity feature close to one of the two footpoints. We show an example in Figure~\ref{event5a}. We point to the main footpoints of the loop with white arrows in the first image and to the corresponding campfire event with a yellow arrow in the last row. The yellow box delimits the negative polarity feature interacting with the positive polarity footpoint.
    
   {\it Category Ib.} In nine cases, the two main footpoints are seen to move towards each other. An example is shown in Figure~\ref{event27b}. The white arrows point to the main footpoints, and the yellow ellipse on the SO/PHI map at 17:11:45 UT marks the location where the negative polarity footpoint splits into a larger and a smaller feature, with the smaller one moving towards the positive polarity footpoint. The yellow arrows in the last row point to the campfire event confined between the two approaching footpoints.

\subsection{Category II: Events with no clear magnetic bipolar features}
In 11 out of 38 campfire events, we cannot clearly link the campfires to two main magnetic footpoints. We assign these campfires to Category II. They either overlie a quiet-Sun region with very weak or little observed $\rm B_{\rm LOS}$, or they occur within a region with randomly scattered small magnetic features. Some of these scattered features are undergoing one or a combination of these processes: merging, splitting, appearance, and disappearance, which is typical for a quiet-Sun region \citep{anusha_2017}. 
If we expand the examined area around some of these events, we find many opposite polarity footpoints; however, it is hard to assign them to clear loops in the EUV images. We show an example of such an event in Figure~\ref{categ2}. The campfire event (yellow circle in both panels) is projected onto a quiet-Sun region, but surrounded by larger magnetic features (white arrows in the upper panel). The spatial distribution of these magnetic features is similar to group (ii) of the categorised magnetic environment around the simulated campfires of \cite{chen_transient_2021}.

\begin{figure*}
\centering
\includegraphics[width=\textwidth]{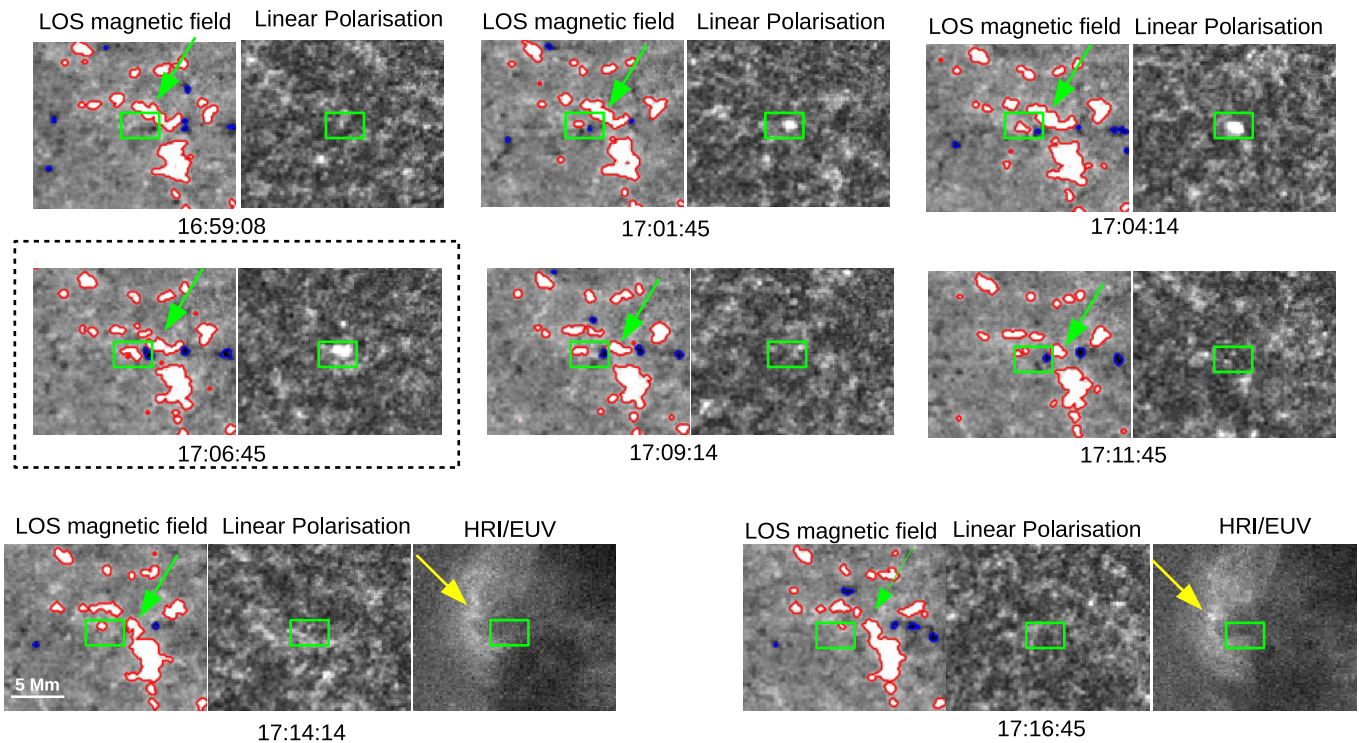}
  \caption{Evolution of the small-scale loop emergence event in the internetwork observed with SO/PHI. The plotted FOV has dimensions of $18\times16$ Mm$^2$. We show in each row the SO/PHI LOS magnetogram (to the left) and the total linear polarisation (to the right). In the last row we show, additionally, the same region in the EUV at the last two timesteps. The green box delimits the two emerging opposite polarity features in the SO/PHI magnetograms and the strong total linear polarisation (TLP) patch in between the emerging magnetic features in the TLP maps. The yellow arrow in the EUV images in the last row points to the campfire event, which has a dot-like shape. The maximum brightness according to the campfire detection algorithm is reached at 17:16:36 UT. The dashed black box at 17:06:45 marks the images at the time of maximum linear polarisation signal and vertical magnetic flux within the green box. 
          }
     \label{event29}
\end{figure*}

\begin{figure}
\centering
\includegraphics[width=\hsize]{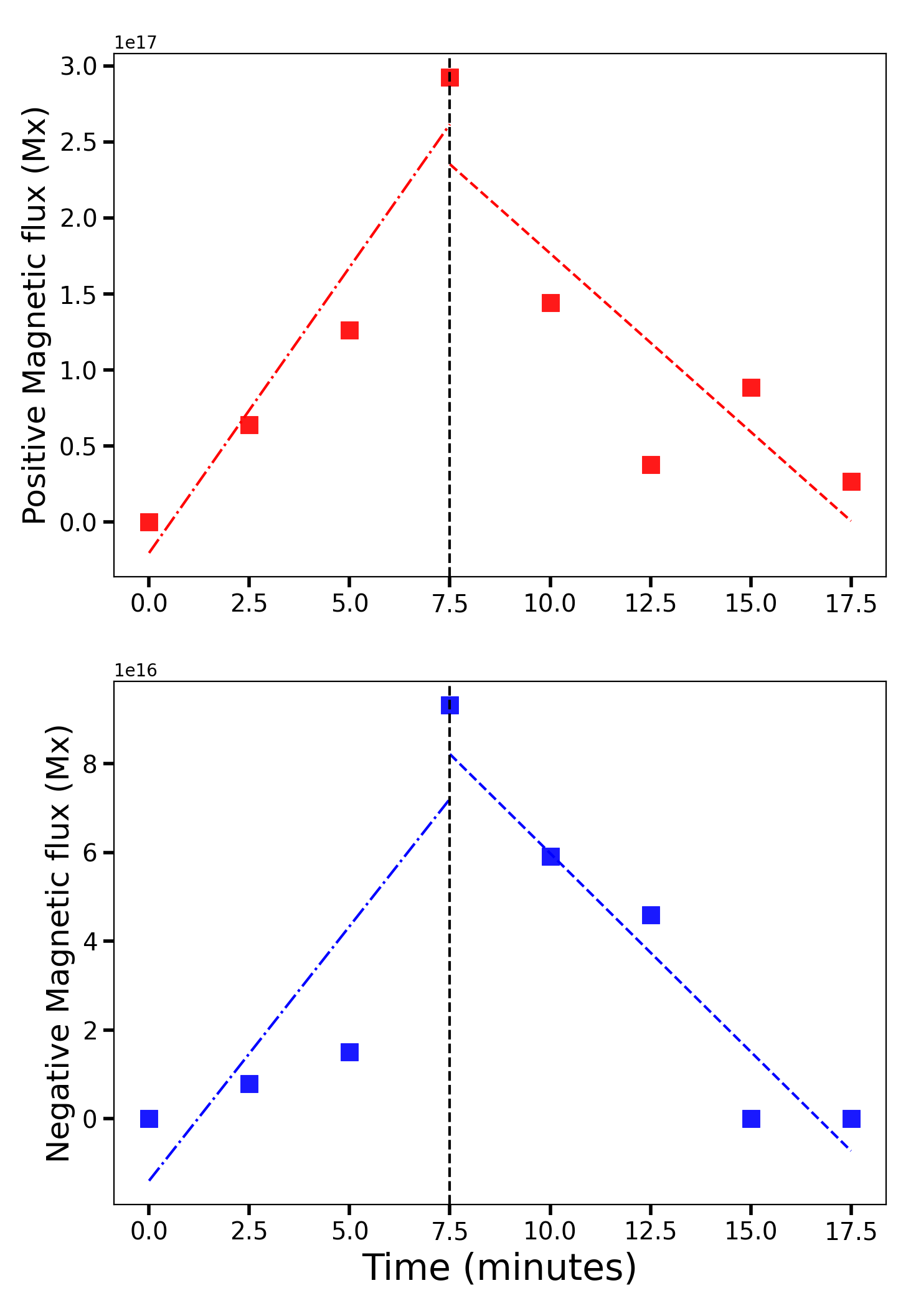}
  \caption{ Magnetic flux variation of the two emerging opposite polarity patches enclosed in a green box in Figure~\ref{event29}. The red and blue dot-dashed lines are the linear fits to the increasing flux for the positive and negative polarities, respectively, while the purely dashed lines are the linear fits to the decreasing part of the flux. The vertical black dashed line corresponds to the time at which the TLP and magnetic flux reach their peak values, thus separating the emergence and cancellation events.
          }
     \label{event29-flux}
\end{figure}
\subsection{Special case: Emergence of a small-scale magnetic loop in the internetwork}
In one particular case, we detected clear signatures of small-scale loop emergence taking place in an internetwork quiet-Sun region. We tracked the emergence event in both the LOS magnetograms and the total linear polarisation maps. We show the detailed evolution of the different magnetic structures at play in Figure~\ref{event29}. The emergence starts with the appearance of two small, opposite polarity footpoints at 17:01:45 UT enclosed in the green box in the $\rm B_{\rm LOS}$ images. The emergence is accompanied by an increase in the magnetic flux of both patches, as indicated by the linear fits to the magnetic flux values of both polarities in Figure~\ref{event29-flux}. The rate of flux emergence for the positive polarity feature is $2\times10^{18}$ Mx/hour and for the negative polarity feature it is $7\times10^{17}$ Mx/hour.

Simultaneously with the emergence of the footpoints in the $\rm B_{\rm LOS}$ maps, a linear polarisation patch (enclosed by the same green box on the TLP images in Figure~\ref{event29}) appears between the two opposite polarity patches. In the continuum image (not shown here), the TLP patch appears inside a granule. The TLP patch strengthens throughout the time series and reaches a maximum value of five times the noise level at a single wavelength ($5\times10^{-3} I_c$) at 17:06:45 (vertical black line in Figure~\ref{event29-flux} and dashed box in Figure~\ref{event29}). After that, the TLP patch starts to disappear while the two footpoints drift apart with the amount of magnetic flux decreasing for both polarities (decreasing trend of the linear fits to the data points after 17:06:45 UT in Figure~\ref{event29-flux}). The rate of flux cancellation is $1.4\times10^{18}$ Mx/hour and $5\times10^{17}$ Mx/hour for the positive and negative polarity features, respectively.

In addition, the negative polarity footpoint emerges close to a larger positive network feature (green arrow on the $\rm B_{\rm LOS}$ maps of Figure~\ref{event29}). At the end of the SO/PHI time series, they undergo cancellation until the negative polarity patch disappears completely in the green box of Figure~\ref{event29} (last two data points with zero flux in the lower panel of Figure~\ref{event29-flux}). The linear polarisation patch in between the footpoints disappears as well.  In the last row, we show the simultaneous images in EUV which display the corresponding campfire event with a dot-like shape (yellow arrow) located adjacent to where the TLP patch was present a few minutes earlier. We could not assign this particular event to Category I since the campfire event is not exactly confined to between the interacting magnetic features in the SO/PHI magnetograms, but we do see signatures of magnetic flux cancellation. Therefore, the event is listed in Category II.

\subsection{Dependence of campfire morphology on magnetic field distribution}
\label{morphology}
The campfire events studied here are either dot-like, loop-like, or events with complex shapes. We followed the categorisation of \cite{berghmans_extreme-uv_2021} of the shape of campfires and checked if there is a correlation between the morphology of campfire events and the magnetic field category. We list the results in Table~\ref{table:1}. For both categories (I and II), the majority of events (55\% in Category I and 54\% in Category II) are mainly loop-like, which is in agreement with \cite{panesar_magnetic_2021}. Point-like events are the second most predominant (30\% in Category I and 36\% in Category II) and events with complex morphologies are seen the least (15\% in Category I and 9\% in Category II). We therefore could not retrieve a clear correlation between the shape of campfires and their respective photospheric magnetic field configuration. 

\begin{table}
\caption{Shape distribution of the campfire events in both categories.}             
\label{table:1}      
\centering                          
\begin{tabular}{c c c c c}        
\hline\hline                 
&Point-like & Loop-like & Complex &Total\\    
\hline                        
  Category I & 8 & 15 & 4&27\\
  \hline
  Category II&4&6&1&11\\
  \hline
  Total & 12&21&5&38\\
\hline                                   
\end{tabular}
\end{table}
\section{Discussion and conclusions}
\label{discussion}

We divided the studied SO/EUI campfire events into two main categories based on the configuration of the SO/PHI photospheric magnetic field below them. The first category is for campfires that are projected onto areas with opposite polarity magnetic features with signatures of magnetic flux cancellation, either between the two footpoints or with a third magnetic feature.

We find similarities in the SO/PHI magnetic configuration at the base of the SO/EUI campfire events of Category I and other UV bursts. For example, the subcategories of the first category have been reported in \cite{mou_magnetic_2016} in a study of coronal bright point formation, using both AIA and HMI data over 84\,hours at a spatial resolution of 1{\arcsec}. In addition, similar to Category Ib (Figure~\ref{event27b}), the scenario of two approaching magnetic bipolar features linked to the formation of CBPs was also reported in \citet{1994ApJ...427..459P,1999ApJ...510L..73P, madjarska_euv_2003, huang_coronal_2012} and modelled in \cite{1994ApJ...427..459P}, for example. The converging motions of two bipolar footpoints that gave rise to Ellerman bombs was also observed by \cite{georgoulis_statistics_2002}.

Magnetic flux cancellation at the site of Category I campfires suggests that they are associated with magnetic reconnection in the lower atmosphere, which could be responsible for their brightening. The same scenario is suggested for the brightening of explosive events \citep{dere_explosive_1992, dere_explosive_1994} and Ellerman bombs  \citep{georgoulis_statistics_2002, rouppe_van_der_voort_reconnection_2016}in the quiet Sun. It has also been proposed for bursts seen in the UV \citep{2014Sci...346C.315P, nelson_relationship_2016}. In addition, magnetic flux cancellation and emergence events were almost always taking place in connection with CBPs \citep[e.g.][]{harvey_does_1999, madjarska_euv_2003, zhang_two_2012, huang_coronal_2012,zhang_reciprocatory_2014, mou_magnetic_2016, madjarska_chromospheric_2021}. 
Cancellation and the associated magnetic reconnection in the lower atmosphere were observed at the footpoints of hotter coronal loops in active regions as well \cite[e.g.][]{chitta_compact_2017, chitta_impulsive_2020}.

We tracked the full emergence event of a small-scale magnetic loop in the internetwork \cite[e.g.][]{centeno_emergence_2007, orozco_suarez_magnetic_2008, martinez_gonzalez_emergence_2009}. The emergence event is seen in both SO/PHI line-of-sight magnetograms and total linear polarisation maps. The emerged footpoints drifted apart, with one of the footpoints interacting later with another existing opposite polarity network feature. Reconnection between the emerged loop and the one originating from the network feature could be the driver to the brightening of the corresponding campfire. The response of the Sun's upper atmosphere to flux emergence events is observed at chromospheric heights \cite[e.g.][]{ortiz_emergence_2014, de_la_cruz_rodriguez_emergence_2015} in AR emergence sites and for transient events in the quiet-Sun corona and coronal holes \cite[e.g.][]{zhang_comparison_2006, huang_coronal_2012}.

The second category of the campfires is for events that could not be linked to clear magnetic footpoints. However, we find in some of these cases opposite polarity features distributed in a larger area around the campfires, the arrangement of which could be similar to the magnetic configuration reported in the simulations of \cite{chen_transient_2021}. This could indicate that they might undergo component reconnection in the corona following the different magnetic field-line configurations described in \cite{chen_transient_2021}. Extrapolations of magnetic field lines using the SO/PHI magnetograms are needed to investigate the topology of the magnetic field lines around these events. 

In addition, a correlation study between the estimated heights of such events and the corresponding photospheric magnetic field configuration can provide insights into the magnetic drivers of such events. Unfortunately, for the HRI$_{\rm EUV}$ dataset studied here, height triangulation with joint observations from Earth is not possible since the observed solar region could not be seen from Earth.
Our current work is based on SO data taken at 0.523 AU from the Sun as well as SO/EUI and SO/PHI time series overlapping for 2.5 minutes. The interesting results obtained with such non-ideal datasets show the potential of the combination of SO/PHI with SO/EUI data obtained during the nominal mission phase. At the perihelion of the orbit of SO, close to 0.3 AU, the higher cadence, higher resolution (200$-$250 km), and longer SO/PHI time series will improve our understanding of the magnetic driving mechanisms of the campfires and how similar they are to other EUV transient events, especially when combined with other remote sensing instruments on board SO such as SPICE \citep{2020A&A...642A..14S}.


\begin{acknowledgements}
We thank the anonymous referee for the constructive comments that helped us improve the manuscript. Solar Orbiter is a space mission of international collaboration between ESA and NASA, operated by ESA. We are grateful to the ESA SOC and MOC teams for their support. The German contribution to SO/PHI is funded by the BMWi through DLR and by MPG central funds. The Spanish contribution is funded by FEDER/AEI/MCIU (RTI2018-096886-C5), a "Center of Excellence Severo
Ochoa" award to IAA-CSIC (SEV-2017-0709), and a Ram\'on y Cajal fellowship awarded to DOS. The French contribution is funded by CNES. The EUI instrument was built by CSL, IAS, MPS, MSSL/UCL, PMOD/WRC, ROB, LCF/IO with funding from the Belgian Federal Science Policy Office (BELSPO/PRODEX PEA 4000112292); the Centre National d'Etudes Spatiales (CNES); the UK Space Agency (UKSA); the Bundesministerium für Wirtschaft und Energie (BMWi) through the Deutsches Zentrum für Luft- und Raumfahrt (DLR); and the Swiss Space Office (SSO).
\end{acknowledgements}

\bibliographystyle{aa}
\bibliography{AA202142873}


\end{document}